\documentclass[aps,preprint,showpacs,preprintnumbers,amsmath,amssymb]{revtex4}
\usepackage{amsmath,mathrsfs,amsbsy,color,graphicx,bm,amsthm,amsfonts}
\usepackage{units}
\usepackage{bbm}
\usepackage{times}
\usepackage{dcolumn}
\usepackage{mathrsfs}% Align table columns on decimal point
\usepackage{amsmath,amssymb,epsfig}
\usepackage{amsmath}
\newcommand{\udots}{\mathinner{\mskip1mu\raise1pt\vbox{\kern7pt\hbox{.}}
\mskip2mu\raise4pt\hbox{.}\mskip2mu\raise7pt\hbox{.}\mskip1mu}}
\begin{document}

\title{Genuine N-partite entanglement and distributed relationships  in the background
of  dilation black holes}
\author{Shu-Min Wu$^1$\footnote{Email: smwu@lnnu.edu.cn}, Yu-Tong Cai$^1$, Wen-Jing Peng$^2$ ,Hao-Sheng Zeng$^2$\footnote{Email: hszeng@hunnu.edu.cn}}
\affiliation{$^1$ Department of Physics, Liaoning Normal University, Dalian 116029, China\\
$^2$ Department of Physics, Hunan Normal University, Changsha 410081, China
}

% \baselineskip=0.65 cm

%\vspace*{0.2cm}
\begin{abstract}
With the complexity of information tasks, the bipartite and tripartite entanglement can no longer meet our needs, and we need more entangled particles to process relativistic quantum information. In this paper, we study the genuine N-partite entanglement and distributed relationships for Dirac fields in the background of dilaton black holes. We present the general analytical
expression including all physically accessible and inaccessible entanglement in curved spacetime.
We find that the accessible N-partite entanglement exhibits irreversible decoherence as the increase of black hole's dilaton, and on the other hand the inaccessible N-partite entanglement increases from zero monotonically or non-monotonically, depending on the relative numbers of the accessible to the inaccessible modes, which forms a sharp contrast with the cases of bipartite and tripartite entanglement where the inaccessible entanglement increase only monotonically. We also find two distributed relationships between  accessible and inaccessible  N-partite entanglement in curved spacetime. The results give us a new understanding of the Hawking radiation.
\end{abstract}

\vspace*{0.5cm}
 \pacs{04.70.Dy, 03.65.Ud,04.62.+v }
\maketitle
\section{Introduction}
The presence of quantum entanglement is widely considered as one of, if not the,
defining property of quantum mechanics \cite{L1}. Since the development of quantum information theory,  entanglement in quantum states has been recognized as a basic resource for quantum computing, quantum cryptography,  quantum teleportation and quantum communication \cite{L2,L3,L4,L5,L6,L7,L8,L9,L10,L11,L12}.
Due to the increasing complexity of quantum information tasks, the bipartite and tripartite entanglement cannot meet the needs. Therefore, we need N-partite entangled state to deal with quantum information tasks. For example, ``Jiuzhang", a quantum computer prototype, which requires $76$ entangled photons, promotes the frontier research of global quantum computing to a new level \cite{L13}. On the other hand, a N-partite quantum state which is not separable with respect to any bipartition is considered  to be genuine N-partite entangled \cite{L4}.
Compared with bipartite entanglement, genuine N-partite entanglement  has obvious advantages in quantum tasks, such as multiparty quantum networks, quantum computing using
cluster states, high sensitivity in metrology tasks and measurement-based
quantum computation \cite{L14,L15,L16,L17,L18,L19,L20,L21,L22,L23}.

The general relativity predicts that black holes exist in our universe. Recently, the first gravitational wave (GW150914) from a binary black hole merger system, was detected by the Advanced LIGO and Virgo detectors \cite{L24}. Moreover, the Event Horizon Telescope revealed the first photo of the supermassive black hole in the center of the giant elliptical galaxy M87 \cite{L25,L26,L27,L28,L29,L30}.
These evidences indirectly and directly confirm the black hole in our universe.
On the other hand, the general theory of relativity and quantum mechanics represent two pillars of modern physics, but unification of the two theories remains an open problem. Quantum gravity tries to solve the contradiction between quantum mechanics and gravity. Some progress has been made in experiments, such as:
(i) gravitational satellite testing  induces quantum decoherence model \cite{L31}; (ii) it is experimentally shown that the gravitational frequency shift effects remarkably influence the precision of atomic clocks for a variation of $0.33 m$ in height \cite{L32}; (iii) the experiment demonstrates direct and full-scale verifications towards ground-satellite quantum key distribution \cite{L33}. In addition, several schemes to simulate quantum gravity have been proposed,
i.e., quantum entanglement in analogue Hawking radiation, analogue Hawking radiation
in a Bose-Einstein Condensate, analogue Hawking radiation and analog cosmological particle generation in a superconducting circuit \cite{L34,L35,L36,L37}.
At the same time,  the bipartite and tripartite entanglement in theory have been widely studied in relativistic framework, such as  noninertial frames and black holes \cite{L38,L39,L40,L41,L42,L43,L44,L45,L46,L47,L48,L49,L50,L51,L52,L53,L54,L55,L56,L57,QLQ57}.
Understanding how the Hawking radiation induced by black holes affects quantum entanglement may help us well understanding the loss of information of black holes \cite{QL57}.

However, N-partite entanglement in relativistic framework has seldom been studied due to the computational complexity. On the other hand, since information tasks are becoming more and more complex, we need more entangled particles to process relativistic quantum information.
It is necessary for us to study N-partite entanglement in curved spacetime. For this purpose, we will discuss properties of genuine N-partite entanglement of Dirac fields in the background of Garfinkle-Horowitz-Strominger (GHS) dilaton black holes. The dilaton black holes
derived from the string theory are formed
by gravitational systems coupled to Maxwell and dilaton fields, and have received a lot of attention \cite{QL1,QL2,QL3,QL4,QL5}.
It is generally believed that  the study of dilaton black holes may deepen the understanding of quantum gravity, as it arises from fundamental theories, such as black hole physics, string theory,  and loop quantum gravity.
In this paper, we study the genuine N-partite entanglement and its distributed relationship for Dirac fields in the background of dilaton black holes. We will obtain the general analytical expression that includes all genuine N-particle entanglement, compare N-particle entanglement with bipartite and tripartite entanglement, and obtain their similar and completely different properties.
In addition, we will look for the distributed relationships between
physically accessible and inaccessible N-partite entanglement, which show the flow law of information  inside and outside the horizon of a black hole.

The paper is organized as follows. In Sec. II, we
discuss the quantization of Dirac fields in the background
of the dilaton black hole. In Sec. III, we
briefly introduce the quantification of genuine N-partite entanglement.
In Sec. IV, we study  the redistribution of genuine N-partite entanglement and  the distributed relationships in dilaton spacetime.  The last section is devoted to a brief conclusion.
%------------------------------------------------------------------------------------------------------------------------------------------------------------------------------------------------%
\section{Quantization of Dirac fields in dilaton spacetime \label{GSCDGE}}
%--------------------------------------------------------------------------------
String theory is a promising candidate for a consistent theory between the general theory of relativity and quantum mechanics. The field according to string theory corresponds to a dilaton with an exponential coupling to an invariant. By choosing the invariant to be the electromagnetic field's Lagrangian, one can get a solution of static dilatonic black hole, i.e., the GHS dilaton black hole \cite{QL1,QL2,QL3,QL4,QL5}. The metric for a GHS dilaton black hole  can be written as \cite{L58,L59}
\begin{eqnarray}\label{Q1}
ds^2=-\left(\frac{r-2M}{r-2D}\right)dt^2+\left(\frac{r-2M}{r-2D}\right)^{-1}
 dr^2+r(r-2D)d
 \Omega^2,
\end{eqnarray}
where $M$ and $D$ are the mass of the black hole and dilaton, respectively. The relationship among the mass $M$, the charge $Q$ and the dilaton $D$ is given by $D=\frac{Q^2}{2M}$.
For simplicity, we take $\hbar, G, c$ and $\kappa_B$ as unity in this paper.

In a general background spacetime, the massless Dirac equation can be expressed as \cite{L60}
\begin{eqnarray}\label{Q2}
[\gamma^a e_a{}^\mu(\partial_\mu+\Gamma_\mu)]\Phi=0,
\end{eqnarray}
where $\gamma^a$ is the Dirac matrices,  the four-vectors $e_a{}^\mu$ is the inverse of the tetrad $e^a{}_\mu$, and $\Gamma_\mu$ is the spin connection coefficient.  Eq.(\ref{Q2})
in the GHS black hole spacetime becomes
\begin{eqnarray}\label{Q3}
&&-\frac{\gamma_0}{\sqrt{f}}\frac{\partial \Phi}{\partial t}+\gamma_1\sqrt{f}\bigg[\frac{\partial}{\partial r}+\frac{r-D}{r \bar r}+\frac{1}{4f}\frac{df}{dr} \bigg]\Phi \nonumber\\
&&+\frac{\gamma_2}{\sqrt{r\bar r}}(\frac{\partial}{\partial \theta}+\frac{\cot \theta}{2})\Phi+\frac{\gamma_3}{\sqrt{r\bar r}\sin\theta}\frac{\partial\Phi}{\partial\varphi}=0,
\end{eqnarray}
where $f=\frac{r-2M}{\bar r}$ with $\bar r=r-2D$.
Then, by solving the Dirac equation in the GHS dilaton black hole, we can obtain the positive
frequency outgoing solutions  for the inside and outside region of the event horizon \cite{L54,L61}
\begin{eqnarray}\label{Q4}
\Phi^+_{{\bold k},{\rm in}}\sim \mathcal{R} e^{i\omega \mathcal{H}},
\end{eqnarray}
\begin{eqnarray}\label{Q5}
\Phi^+_{{\bold k},{\rm out}}\sim \mathcal{R} e^{-i\omega \mathcal{H}},
\end{eqnarray}
where $\mathcal{R}$ is a four-component Dirac spinor, $\mathcal{H}=t-r_{*}$ is the tortoise coordinate and $\bold k$ is the wave vector labeling the modes hereafter.
Eqs.(\ref{Q4}) and (\ref{Q5}) can be used to expand the Dirac field as
\begin{eqnarray}\label{Q6}
\Phi=\sum_\sigma\int
d\bold k[\hat{a}^{\sigma}_{\bold k}\Phi^{+}_{{\bold k},\sigma}
+\hat{b}^{\sigma\dag}_{\bold k}
\Phi^{-}_{{\bold k},\sigma}],
\end{eqnarray}
where $\sigma=(\rm in, \rm out)$, $\hat{a}^{\sigma}_{\bold k}$ and $\hat{b}^{\sigma\dag}_{\bold k}$ are the fermion annihilation and antifermion creation operators acting on  the quantum state, respectively. The annihilation and creation operators  satisfy the canonical anticommutation
$$\{\hat{a}^{\rm in}_{\mathbf{k}},\hat{a}^{\rm in\dagger}_{\mathbf{k'}}\}=\{\hat{a}^{\rm out}_{\mathbf{k}},\hat{a}^{\rm out\dagger}_{\mathbf{k'}}\}=
\{\hat{b}^{\rm in}_{\mathbf{k}},\hat{b}^{\rm in\dagger}_{\mathbf{k'}}\}=\{\hat{b}^{\rm out}_{\mathbf{k}},\hat{b}^{\rm out\dagger}_{\mathbf{k'}}\}
=\delta_{\mathbf{k}\mathbf{k'}}. $$
One  can  define the dilaton vacuum $\hat{a}^{in}_{\bold k}|0\rangle_D=\hat{a}^{out}_{\bold k}|0\rangle_D=0$.
Therefore, the
modes  $\Phi^\pm_{{\bold k},{\sigma}}$ and $\Phi^\pm_{{\bold k},{\sigma}}$ are called dilaton modes.

Making analytic continuations for Eqs.(\ref{Q4}) and (\ref{Q5}), according to the suggestion of
Domour-Ruffini \cite{L62}, we find a complete basis for positive energy modes,
i.e., the Kruskal modes. Then we can quantize the
Dirac field in Kruskal modes
\begin{eqnarray}\label{Q7}
\Phi=\sum_\sigma\int
d\bold k \frac{1}{\sqrt{[2\cosh(4\pi(M-D)\omega)]}}
[\hat{c}^{ \sigma}_{\bold k}\Psi^{+}_{{\bold k},\sigma}
+\hat{d}^{\sigma\dag}_{\bold k}
\Psi^{-}_{{\bold k},\sigma}],
\end{eqnarray}
where $\hat{c}^{\sigma}_{\bold k}$ and $\hat{d}^{\sigma\dag}_{\bold k}$ are the fermion annihilation and antifermion creation operators acting on the Kruskal vacuum.
Eqs.(\ref{Q6}) and (\ref{Q7}) show that the Dirac field is decomposed into
the dilaton and Kruskal modes, respectively. Now, we can easily obtain the Bogoliubov
transformations  between annihilation and creation
operators in Kruskal and dilaton coordinates.
Using the Bogoliubov transformations,
the relations between the Kruskal and dilaton operators take the forms
\begin{eqnarray}\label{Q8}
\hat{c}^{\rm out}_{\bold k}&=&\alpha\hat{a}^{\rm out}_{\bold k}-\beta\hat{b}^{\rm out\dag}_{\bold k},\\ \nonumber
\hat{c}^{\rm out\dag}_{\bold k}&=&\alpha\hat{a}^{\rm out\dag}_{\bold k}-\beta\hat{b}^{\rm out}_{\bold k},
\end{eqnarray}
with $\alpha=\frac{1}{\sqrt{e^{-8\pi (M-D)\omega}+1}}$ and $\beta=\frac{1}{\sqrt{e^{8\pi (M-D)\omega}+1}}$.
Since the dilaton black hole can be divided to the physically accessible and inaccessible  region, the ground and only excited states in Kruskal coordinate correspond to a two-mode squeezed state in  dilaton black hole coordinate. After properly normalizing the state vector, the Kruskal vacuum state and the only excited state can be given by
\begin{eqnarray}\label{Q9}
\nonumber |0\rangle_K&=&\alpha|0\rangle_{\rm out} |0\rangle_{\rm in}+\beta|1\rangle_{\rm out} |1\rangle_{\rm in},\\
|1\rangle_K&=&|1\rangle_{\rm out} |0\rangle_{\rm in},
\end{eqnarray}
where  $\{|n\rangle_{\rm out}\}$ and $\{|n\rangle_{\rm in}\}$ are the number states for the fermion outside the region and the antifermion inside the region of the event horizon, respectively.
%------------------------------------------------------------------------------------------------------------------------------------------------------------------------------------------------%
\section{ Quantification of genuine N-partite entanglement \label{GSCDGE}}
%--------------------------------------------------------------------------------
The genuine N-partite entanglement is an essential resource for quantum computing and quantum networks. Next, we briefly review the measure of genuine N-partite entanglement.
Generally, N-partite entanglement is defined by its opposite of biseparability. A biseparable N-partite pure state $|\Psi\rangle$ can be written as $|\Psi\rangle=|\Psi_A\rangle\otimes|\Psi_B\rangle$, with
$|\Psi_A\rangle\in H_A=H_{i_1}\otimes H_{i_2}\otimes...\otimes H_{i_k}$ and
$|\Psi_B\rangle\in H_B=H_{i_{k+1}}\otimes H_{i_{k+2}}\otimes...\otimes H_{i_N}$ in any  bipartition of the Hilbert space. A biseparable  N-partite mixed state $\rho$  is given by a convex combination of biseparable states $\rho=\sum_ip_i|\Psi_i\rangle\langle\Psi_i|$,
where the $|\Psi_i\rangle$ can be biseparable with respect to different bipartitions.
If an N-partite  entangled state is not biseparable, it is called genuinely N-partite  entangled state. In other words, the genuine N-partite entanglement is not separable with respect to any bipartition. The genuine  N-partite entanglement for pure state is defined as \cite{L63}
\begin{eqnarray}\label{Q10}
E(|\Psi\rangle)=\min_{\chi_i\in\chi}\sqrt{2[1-\text{Tr}(\rho^2_{A_{i}})]},
\end{eqnarray}
where $\chi=\{A_i|B_i\}$ denotes the set of all possible bipartitions of the whole $N$-partite system, and $\rho_{A_{i}}$ is the marginal state for the subsystem $A_{i}$.
The genuine N-partite entanglement for mixed states can be obtained via a convex roof construction
\begin{eqnarray}\label{Q11}
E(\rho)=\inf_{\{p_i,|\Psi_i\rangle\}}\sum_ip_iE(|\Psi_i\rangle),
\end{eqnarray}
where the infimum is taken over all possible decompositions
$\rho=\sum_ip_i|\Psi_i\rangle\langle\Psi_i|$.

For the N-qubit systems, the Hilbert-space orthonormal bases are usually defined as $\{|0,0,...,0\rangle,|0,0,...,1\rangle,...,|1,1,...,1\rangle\}$. The X-state of a N-qubit system may be written in terms of density matrix as
\begin{eqnarray}\label{Q12}
 \rho_X= \left(\!\!\begin{array}{cccccccc}
a_1 &  &  &  &  &  &  & c_1\\
 & a_2 &  &  &  &  & c_2 & \\
 &  & \ddots &  &  &  \udots &  & \\
 &  &  & a_n & c_n &  &  & \\
 &  &  & c_n^* &  b_n & &  & \\
 &  & \udots &  &  & \ddots &  & \\
 & c_2^* &  &  &  &  & b_2 & \\
c_1^* &  &  &  &  &  &  & b_1
\end{array}\!\!\right),
\end{eqnarray}
where $n=2^{N-1}$.  The conditions $\sum_i(a_i+b_i)=1$ and $|c_i|\leq\sqrt{a_ib_i}$
are required to ensure that $\rho_X$ is normalized and positive.
 The genuine entanglement for a N-qubit X-state can be expressed as
\begin{eqnarray}\label{Q13}
E(\rho_X)=2\max \{0,|c_i|-\nu_i \},   i=1,\ldots,n,
\end{eqnarray}
where $\nu_i=\sum_{j\neq i}^n\sqrt{a_jb_j}$ \cite{L64}.

%------------------------------------------------------------------------------------------------------------------------------------------------------------------------------------------------%
\section{ Genuine N-partite entanglement of Dirac fields in GHS dilaton black hole  \label{GSCDGE}}
%--------------------------------------------------------------------------------
We  initially share an N-partite  entangled state in the asymptotically flat region of the
dilaton black hole
\begin{eqnarray}\label{Q14}
|\psi\rangle_{1,\ldots,N}=\cos\theta|0\rangle^{\otimes N}+\sin\theta|1\rangle^{\otimes N},
\end{eqnarray}
where $\theta\in(0,\frac{\pi}{2})$ and the mode $i$ ($i = 1, 2, \ldots, N$) is observed by observer $O_i$.
We assume that $n$ ($n<N$) observers hover near the event horizon of the GHS
dilaton black hole, and the rest $N-n$ observers stay stationarily at the asymptotically flat region. Using Eq.(\ref{Q9}), we can rewrite Eq.(\ref{Q14}) in terms of dilaton modes
\begin{eqnarray}\label{Q15}
|\psi\rangle_{1,\ldots,N+n}&=&\cos\theta\left[(\overbrace{|0\rangle_{n+1}|0\rangle_{n+2}...|0\rangle_{N}}^{|\bar 0\rangle})\bigotimes_{i=1}^{n}(\alpha|0\rangle_{{\rm out,}i} |0\rangle_{{\rm in,}i}
+\beta|1\rangle_{{\rm out,}i} |1\rangle_{{\rm in,}i})\right]\\
\nonumber &+&\sin\theta\left[(\overbrace{|1\rangle_{n+1}|1\rangle_{n+2}...|1\rangle_{N}}^{|\bar 1\rangle})\bigotimes_{i=1}^{n}
(|1\rangle_{{\rm out,}i} |0\rangle_{{\rm in,}i})\right],
\end{eqnarray}
where $|\bar 0\rangle=|0\rangle_{n+1}|0\rangle_{n+2}...|0\rangle_{N}\rangle$
and $|\bar 1\rangle=|1\rangle_{n+1}|1\rangle_{n+2}...|1\rangle_{N}\rangle$.
Since the exterior region is causally disconnected from the interior region of the black hole, we should trace over the state of the inside region and obtain
\begin{eqnarray}\label{Q16}
\rho_{1,\ldots,N}^{\rm out}&=&\alpha^{2n}\cos^2\theta|\bar0 \rangle\langle\bar0|\bigotimes_{i=1}^{n}(|0\rangle_{{\rm out,}i}\langle0|)\\
\nonumber &+&\alpha^{2n-2}\beta^{2}\cos^2\theta|\bar0 \rangle\langle\bar0|\sum_{m=1}^n \left[(|1\rangle_{{\rm out,}m}\langle1|)\bigotimes_{i=1,i\neq m}^{n}(|0\rangle_{{\rm out,}i}\langle0|)\right]\\
\nonumber &+&\alpha^{2n-4}\beta^{4}\cos^2\theta|\bar0 \rangle\langle\bar0|\sum_{m=2}^n\sum_{z=1}^{m-1}\left[(|1\rangle_{{\rm out,}z}\langle1|)(|1\rangle_{{\rm out,}m}\langle1|)\bigotimes_{i=1,i\neq z,m}^{n}(|0\rangle_{{\rm out,}i}\langle0|)\right]\\
\nonumber&+&...\\
\nonumber &+&\beta^{2n}\cos^2\theta|\bar0 \rangle\langle\bar0|\bigotimes_{i=1}^{n}(|1\rangle_{{\rm out,}i}\langle1|)\\
\nonumber &+&\alpha^{n}\cos\theta\sin\theta\bigg\{[|\bar0 \rangle\langle\bar1|\bigotimes_{i=1}^{n}(|0\rangle_{{\rm out,}i}\langle1|)]+h.c.\bigg\}\\
\nonumber &+&\sin^2\theta|\bar1 \rangle\langle\bar1|\bigotimes_{i=1}^{n}(|1\rangle_{{\rm out,}i}\langle1|), \nonumber
\end{eqnarray}
which we write in matrix form as
\begin{eqnarray}\label{Q17}
 \rho_{1,\ldots,N}^{\rm out}= \left(\!\!\begin{array}{cc}
 \mathcal{A}_{\rm out} & \mathcal{C}_{\rm out} \\
 \mathcal{C}^T_{\rm out} & \mathcal{B}_{\rm out} \\
 \end{array}\!\!\right),
\end{eqnarray}
in the $2^{n+1}$ basis $\{|\bar00...00\rangle,|\bar00...01\rangle,...,|\bar11...10\rangle,|\bar11...11\rangle \}.$
The sub-matrixes $\mathcal{A}_{\rm out}$, $\mathcal{B}_{\rm out}$, $\mathcal{C}_{\rm out}$ and $\mathcal{C}_{\rm out}^T$ are $2^n\times2^n$ dimensions.
The basis set for sub-matrixes $\mathcal{A}_{\rm out}$ is $\{|\bar00...00\rangle, \ldots, |\bar01...11\rangle\}$, where the base corresponding to the element $\alpha^{2n-2i}\beta^{2i}\cos^2\theta$ include $i$ ``1" .
The sub-matrixes can be written as
\begin{eqnarray}\label{QQ24}
\mathcal{A}_{\rm out}=\cos^2\theta \left(\!\!\begin{array}{cccc}
\alpha^{2n}\mathbf{I}_{0} &  & &\\
  & \alpha^{2n-2}\beta^{2}\mathbf{I}_{1} & &\\
   &  & \ddots&\\
  &  & &\beta^{2n}\mathbf{I}_{n} \\
 \end{array}\!\!\right),
\end{eqnarray}
\begin{eqnarray}\label{QQ244}
\mathcal{B}_{\rm out}= \left(\!\!\begin{array}{cccc}
0 &  & &\\
  & \ddots & &\\
   &  & 0&\\
  &  & &\sin^2\theta \\
 \end{array}\!\!\right),
\end{eqnarray}
\begin{eqnarray}\label{QQ25}
\mathcal{C}_{\rm out}= \left(\!\!\begin{array}{cccc}
 &  & &\alpha^{n}\cos\theta\sin\theta\\
  &  &0 &\\
   & \udots& &\\
 0 &  & &\\
 \end{array}\!\!\right),
\end{eqnarray}
where the $\mathbf{I}_{i}$ in (\ref{QQ24}) denotes a $C_{n}^{i}\times C_{n}^{i} $ identity matrix with $C_{n}^{i}=\frac{n!}{i!(n-i)!}$ the binomial coefficient.

According to Eqs.(\ref{Q12}) and (\ref{Q17}), we can obtain the  physically accessible genuine N-partite entanglement
\begin{eqnarray}\label{Q18}
E(\rho_{1,\ldots,N}^{\rm out})=2\max \bigg\{0,\alpha^{n}\cos\theta\sin\theta \bigg\}.
\end{eqnarray}
From Eq.(\ref{Q18}), we can see that the genuine N-partite entanglement $E(\rho_{1,\ldots,N}^{\rm out})$ depends not only on the initial parameter $\theta$, but also on the dilaton $D$ of the black hole; this means that the effect of gravitation induced by the black hole will affect the entanglement. We can also see that the genuine N-partite entanglement $E(\rho_{1,\ldots,N}^{\rm out})$ is independent of N, but depends on $n$. In addition, we obtain $\frac{\partial E(\rho_{1,\ldots,N}^{\rm out})}{\partial \theta}=2\alpha^{n}\cos2\theta$, which means that, for the given  dilaton $D$ and $n$,  the maximum of genuine N-partite entanglement $E(\rho_{1,\ldots,N}^{\rm out})$ corresponds to $\theta=\frac{\pi}{4}$. This result is different from tripartite tangles, where the maximum takes place at different initial parameters \cite{L56}.

\begin{figure}
\begin{minipage}[t]{0.5\linewidth}
\centering
\includegraphics[width=3.0in,height=5.2cm]{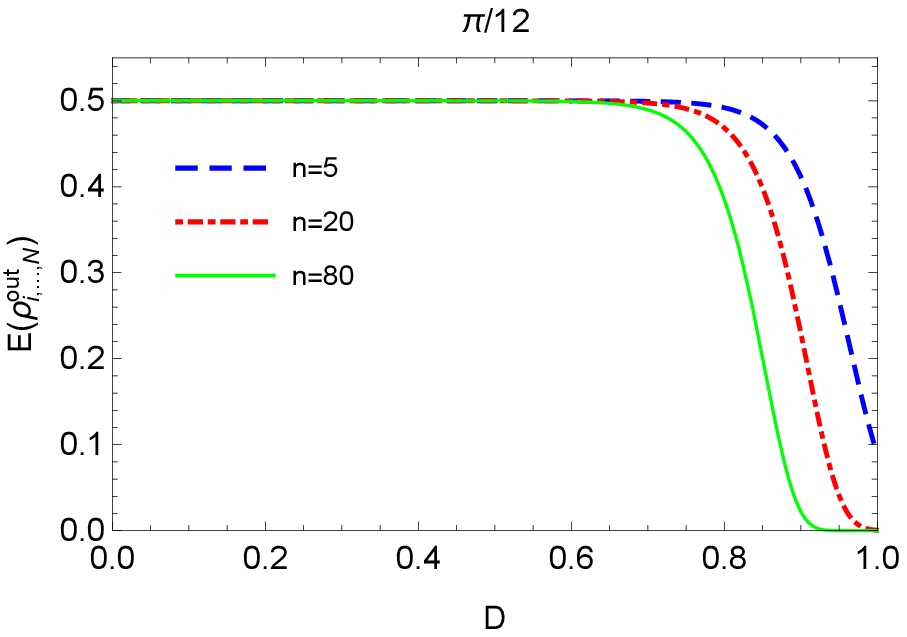}
\label{fig1a}
\end{minipage}%
\begin{minipage}[t]{0.5\linewidth}
\centering
\includegraphics[width=3.0in,height=5.2cm]{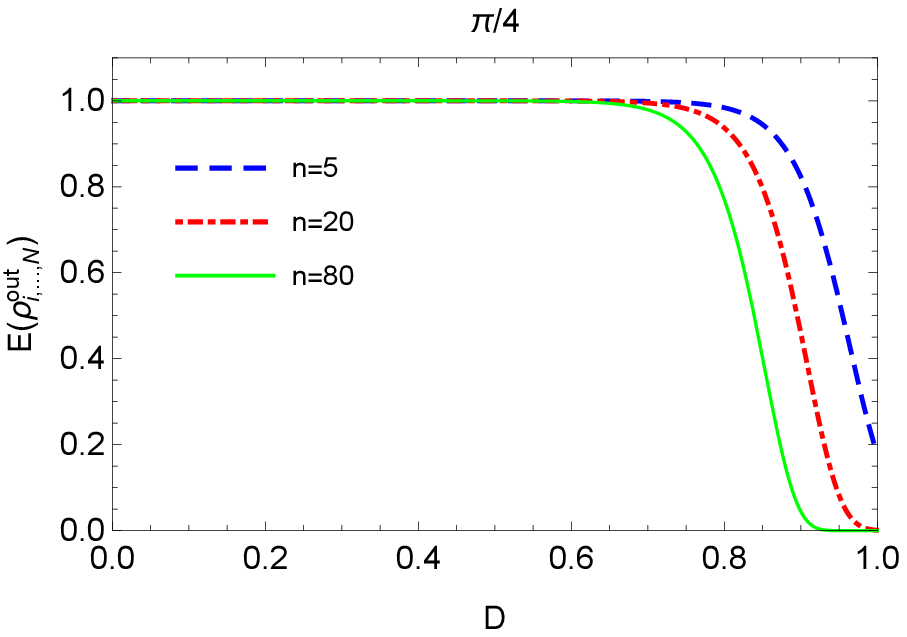}
\label{fig1c}
\end{minipage}%
\caption{The genuine N-partite entanglement $E(\rho_{1,\ldots,N}^{\rm out})$ as functions of the  dilaton $D$ of the GHS black hole for different $n$ and $\theta$, where $M=\omega=1$. $n=5$ (dashed blue line); $n=20$ (dotted red line).; $n=80$ (solid green line).}
\label{Fig1}
\end{figure}

In Fig.\ref{Fig1}, we show the genuine N-partite entanglement $E(\rho_{1,\ldots,N}^{\rm out})$ as functions of the dilaton $D$ for different $n$ and $\theta$.
We find  that the entanglement $E(\rho_{1,\ldots,N}^{\rm out})$ decreases with the increase of the  dilaton $D$, meaning that the gravitational effect  degrades the physically accessible quantum entanglement.
For the extreme black hole ($D\rightarrow M$), we find from Eq.(\ref{Q18}) that
$$\lim_{D\rightarrow M}E(\rho_{1,\ldots,N}^{\rm out})=\sin2\theta(\frac{1}{\sqrt{2}})^n.$$
Fig.\ref{Fig1} also shows that, for given dilaton $D$, $E(\rho_{1,\ldots,N}^{\rm out})$ reduces monotonically with $n$. This is because Hawking radiation behaves like a kind of noise to the entanglement. The larger the $n$ is, the more the modes influenced by Hawking radiation are, and thus the more $E(\rho_{1,\ldots,N}^{\rm out})$ declines. Especially for the extreme black hole ($D\rightarrow M$), and in the limit of infinite $n$, the entanglement $E(\rho_{1,\ldots,N}^{\rm out})$ vanishes.

Above, we have discussed the genuine N-partite entanglement for the modes: $N-n$ modes in the asymptotically flat region and $n$ modes outside the event horizon. We call this kind of entanglement is physically inaccessible. Now we discuss another N-partite entanglement which consists of $N-n$ modes in the asymptotically flat region and $n$ modes inside the event horizon. This N-partite entanglement is called physically inaccessible. By tracing over physically accessible $n$ modes outside the horizon in Eq.(\ref{Q15}), we obtain the density operator
\begin{eqnarray}\label{Q19}
\rho_{1,\ldots,N}^{\rm in}&=&\alpha^{2n}\cos^2\theta|\bar0 \rangle\langle\bar0|\bigotimes_{i=1}^{n}(|0\rangle_{{\rm in,}i}\langle0|)\\ \nonumber
&+&\alpha^{2n-2}\beta^{2}\cos^2\theta|\bar0 \rangle\langle\bar0|\sum_{m=1}^n \left[(|1\rangle_{{\rm in,}m}\langle1|)\bigotimes_{i=1,i\neq m}^{n}(|0\rangle_{{\rm in,}i}\langle0|)\right]\\ \nonumber
&+&\alpha^{2n-4}\beta^{4}\cos^2\theta|\bar0 \rangle\langle\bar0|\sum_{m=2}^{n}\sum_{z=1}^{m-1}\left[(|1\rangle_{{\rm in,}z}\langle1|)(|1\rangle_{{\rm in,}m}\langle1|)\bigotimes_{i=1,i\neq z,m}^{n}(|0\rangle_{{\rm in,}i}\langle0|)\right]
\\ \nonumber &+&...
\\ \nonumber
&+&\beta^{2n}\cos^2\theta|\bar0 \rangle\langle\bar0|\bigotimes_{i=1}^{n}(|1\rangle_{{\rm in,}i}\langle1|)\\ \nonumber
&+&\beta^{n}\cos\theta\sin\theta\bigg\{[|\bar0 \rangle\langle\bar1|\bigotimes_{i=1}^{n}(|1\rangle_{{\rm in,}i}\langle0|)]+h.c. \bigg\}
\\ \nonumber &+&\sin^2\theta|\bar1 \rangle\langle\bar1|\bigotimes_{i=1}^{n}(|0\rangle_{{\rm in,}i}\langle0|),
\end{eqnarray}
which we write in matrix form as
\begin{eqnarray}\label{Q20}
 \rho_{1,\ldots,N}^{\rm in}= \left(\!\!\begin{array}{cc}
 \mathcal{A}_{\rm in} & \mathcal{C}_{\rm in} \\
 \mathcal{C}^T_{\rm in} & \mathcal{B}_{\rm in} \\
 \end{array}\!\!\right).
\end{eqnarray}
where $\mathcal{A}_{\rm in}=\mathcal{A}_{\rm out}$,
\begin{eqnarray}\label{QQq24}
\mathcal{B}_{\rm in}= \left(\!\!\begin{array}{cccc}
\sin^2\theta &  & &\\
  & 0 & &\\
   &  & \ddots&\\
  &  & &0 \\
 \end{array}\!\!\right),
\end{eqnarray}
and
\begin{eqnarray}\label{QQq25}
\mathcal{C}_{\rm in}= \left(\!\!\begin{array}{cccc}
 &  & &0\\
  &  &\udots &\\
   & 0& &\\
\beta^{n}\cos\theta\sin\theta &  & &\\
 \end{array}\!\!\right).
\end{eqnarray}

In a similar way, we calculate the genuine N-partite entanglement for the  physically inaccessible modes
\begin{eqnarray}\label{Q21}
E(\rho_{1,\ldots,N}^{\rm in})=2\max \bigg\{0,\beta^{n}\cos\theta\sin\theta \bigg\}.
\end{eqnarray}

\begin{figure}
\begin{minipage}[t]{0.5\linewidth}
\centering
\includegraphics[width=3.0in,height=5.2cm]{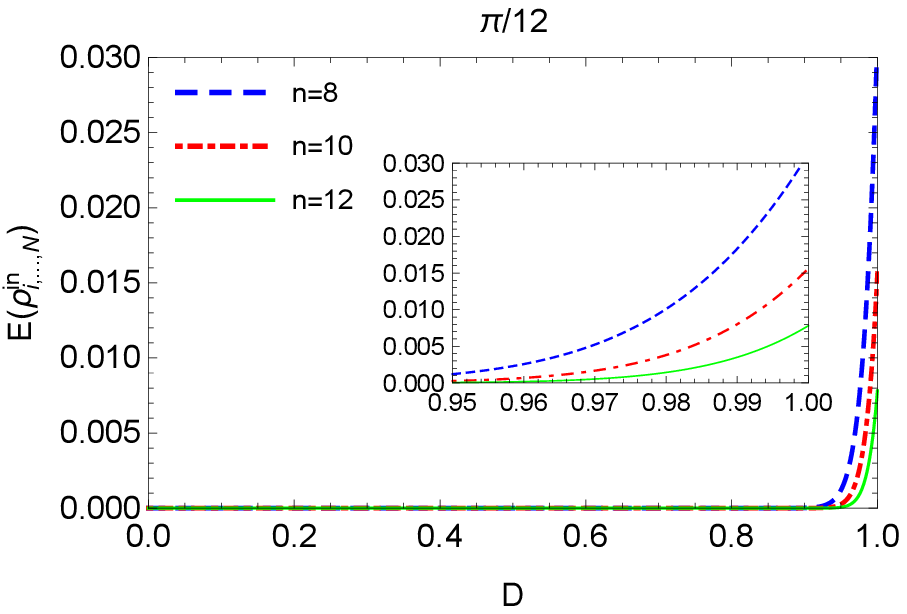}
\label{fig1a}
\end{minipage}%
\begin{minipage}[t]{0.5\linewidth}
\centering
\includegraphics[width=3.0in,height=5.2cm]{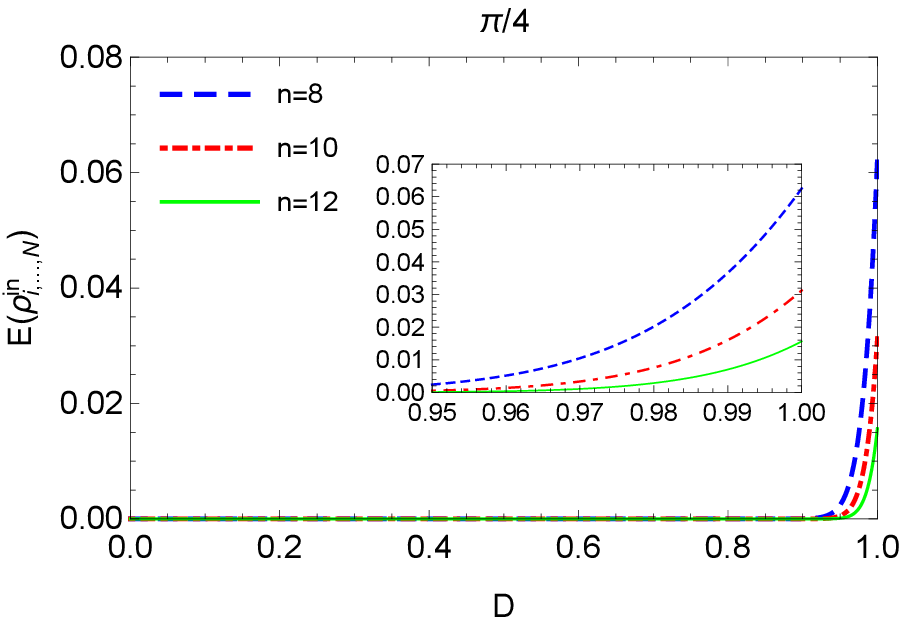}
\label{fig1c}
\end{minipage}%
\caption{The genuine N-partite entanglement $E(\rho_{1,\ldots,N}^{\rm in})$ as functions of the  dilaton $D$ of the GHS black hole for different $n$ and $\theta$, where $M=\omega=1$. $n=8$ (dashed blue line); $n=10$ (dotted red line).; $n=12$ (solid green line).}
\label{Fig2}
\end{figure}

Fig.\ref{Fig2} shows how the dilaton $D$ of the black hole
influences the genuine N-partite entanglement $E(\rho_{1,\ldots,N}^{\rm in})$ between the  physically inaccessible modes. From which we can see that the  entanglement $E(\rho_{1,\ldots,N}^{\rm in})$ increases with the growth of the dilaton $D$,
which means that the gravitational effect produced by the black hole can generate the physically inaccessible entanglement.
For the extreme black hole, we obtain from the analytic expression of Eq.(\ref{Q21}) that $$\lim_{D\rightarrow M}E(\rho_{1,\ldots,N}^{\rm in})=\sin2\theta(\frac{1}{\sqrt{2}})^n.$$
We also see from Fig.\ref{Fig2} that, for the given dilaton $D$, the  $E(\rho_{1,\ldots,N}^{\rm in})$ decreases monotonically with the increase of the $n$. We can understand this question as follows: Hawking radiation causes information traveling from the exterior to the interior of the event horizon, leading to the generation of the entanglement $E(\rho_{1,\ldots,N}^{\rm in})$. In this process, the event horizon of black hole plays the role of resistance. The larger the $n$ is, the stronger the resistance is, and thus the slower $E(\rho_{1,\ldots,N}^{\rm in})$ rises.

So far, we have studied the genuine N-partite entanglement for the accessible and inaccessible modes in curved spacetime.  Next, we extend it to a more general case. For this end,
we consider a general system $\rho_{N-n,p,q}$ ($p+q=n$), which consists of $N-n$  modes in the asymptotically flat region, physically accessible $p$ modes outside the event horizon and  physically inaccessible $q$ modes inside the event horizon.
The density operator $\rho_{N-n,p,q}$  can be expressed as
\begin{eqnarray}\label{Q22}
\rho_{N-n,p,q}&=&\alpha^{2n}\cos^2\theta|\bar0 \rangle\langle\bar0|\left[\bigotimes_{i=1}^{p}(|0\rangle_{{\rm out,}i}\langle0|)\right]\left[\bigotimes_{j=1}^{q}(|0\rangle_{{\rm in,}j}\langle0|)\right]\\ \nonumber
&+&\alpha^{2n-2}\beta^{2}\cos^2\theta|\bar0 \rangle\langle\bar0|\bigg\{\sum_{m=1}^p [(|1\rangle_{{\rm out,}m}\langle1|)\bigotimes_{i=1,i\neq m}^{p}(|0\rangle_{{\rm out,}i}\langle0|)\bigotimes_{j=1}^{q}(|0\rangle_{{\rm in,}j}\langle0|)]\\ \nonumber
&&\hspace{0.3cm} +\sum_{m=1}^q [\bigotimes_{i=1}^{p}(|0\rangle_{{\rm out,}i}\langle0|)(|1\rangle_{{\rm in,}m}\langle1|)\bigotimes_{j=1,j\neq m}^{q}(|0\rangle_{{\rm in,}j}\langle0|)]\bigg\}\\ \nonumber
&+&\alpha^{2n-4}\beta^{4}\cos^2\theta|\bar0 \rangle\langle\bar0|
\\ \nonumber && \hspace{0.3cm}\times\bigg\{\sum_{m=2}^p\sum_{z=1}^{m-1} [(|1\rangle_{{\rm out,}z}\langle 1|)(|1\rangle_{{\rm out,}m}\langle 1|)\bigotimes_{i=1,i\neq z,m}^{p}(|0\rangle_{{\rm out,}i}\langle0|)\bigotimes_{j=1}^{q}(|0\rangle_{{\rm in,}j}\langle0|)]
\\ \nonumber && \hspace{0.3cm}+\sum_{m=2}^q \sum_{z=1}^{m-1}[\bigotimes_{i=1}^{p}(|0\rangle_{{\rm out,}i}\langle0|)(|1\rangle_{{\rm in,}z}\langle1|)(|1\rangle_{{\rm in,}m}\langle1|)\bigotimes_{j=1,j\neq z,m}^{q}(|0\rangle_{{\rm in,}j}\langle0|)]
\\ \nonumber && \hspace{0.3cm}+\sum_{z=1}^p\sum_{m=1}^q [(|1\rangle_{{\rm out,}z}\langle1|)\bigotimes_{i=1,i\neq z}^{p}(|0\rangle_{{\rm out,}i}\langle0|)(|1\rangle_{{\rm in,}m}\langle1|)\bigotimes_{j=1,j\neq m}^{q}(|0\rangle_{{\rm in,}j}\langle0|)]\bigg\}\\ \nonumber
&+&...\\ \nonumber
&+&\beta^{2n}\cos^2\theta|\bar0 \rangle\langle\bar0|\bigotimes_{i=1}^{p}(|1\rangle_{{\rm out,}i}\langle1|)\bigotimes_{j=1}^{q}(|1\rangle_{{\rm in,}j}\langle1|)\\ \nonumber
&+&\alpha^{p}\beta^{q}\cos\theta\sin\theta\bigg\{|\bar0 \rangle\langle\bar1|\bigotimes_{i=1}^{p}(|0\rangle_{{\rm out,}i}\langle1|)\bigotimes_{j=1}^{q}(|1\rangle_{{\rm in,}j}\langle0|)+h.c.\bigg\}\\ \nonumber
&+&\sin^2\theta|\bar1 \rangle\langle\bar1|\bigotimes_{i=1}^{p}(|1\rangle_{{\rm out,}i}\langle1|)\bigotimes_{j=1}^{q}(|0\rangle_{{\rm in,}j}\langle0|).
\end{eqnarray}
The density operator $\rho_{N-n,p,q}$ has the matrix form
\begin{eqnarray}\label{Q23}
 \rho_{N-n,p,q}= \left(\!\!\begin{array}{cc}
 \mathcal{A} & \mathcal{C} \\
 \mathcal{C}^T & \mathcal{B} \\
 \end{array}\!\!\right),
\end{eqnarray}
where $\mathcal{A}=\mathcal{A}_{\rm out}$.
The sub-matrixes $\mathcal{A}$, $\mathcal{B}$ and $\mathcal{C}$ are $2^n\times2^n$ dimensions. Here, the sub-matrixes $\mathcal{B}$ and $\mathcal{C}$ are given by
\begin{eqnarray}\label{Q25}
\mathcal{B}= \left(\!\!\begin{array}{ccccccc}
 0 &  & & & & & \\
  &  \ddots& & & & & \\
  &  & 0& & & & \\
  &  & & \sin^2\theta& & & \\
  &  & & &0 & & \\
  &  & & & &\ddots & \\
  &  & & & & & 0\\
 \end{array}\!\!\right),
\end{eqnarray}
where  the element $\sin^2\theta$ is fixed in the position ($2^{\rm p},2^{\rm p}$), and
\begin{eqnarray}\label{Q26}
\mathcal{C}= \left(\!\!\begin{array}{ccccccc}
  &  & & & & &0 \\
  &  & & & & \udots& \\
  &  & & & 0& & \\
  &  & & \alpha^{p}\beta^{q}\cos\theta\sin\theta & & & \\
  &  & 0& & & & \\
  &  \udots& & & & & \\
  0&  & & & & & \\
 \end{array}\!\!\right),
\end{eqnarray}
where the element $\alpha^{p}\beta^{q}\cos\theta\sin\theta$ is fixed in the position ($2^{\rm q},2^{\rm p}$).
Substitution of the elements of the matrix $\rho_{N-n,p,q}$ into Eq.(\ref{Q12}), we obtain
\begin{eqnarray}\label{Q27}
E(\rho_{N-n,p,q})=2\max \bigg\{0,\alpha^{p}\beta^{q}\cos\theta\sin\theta \bigg\}.
\end{eqnarray}
We note that: (i) for $p=n$, $E(\rho_{N-n,p,q})$  changes to $E(\rho_{1,\ldots,N}^{\rm out})$; (ii) for $q=n$, we obtain $E(\rho_{N-n,p,q})=E(\rho_{1,\ldots,N}^{\rm in})$.
Therefore, $E(\rho_{N-n,p,q})$ is a more general expression which includes both $E(\rho_{1,\ldots,N}^{\rm out})$ and $E(\rho_{1,\ldots,N}^{\rm in})$ as the special cases.

\begin{figure}
\begin{minipage}[t]{0.5\linewidth}
\centering
\includegraphics[width=3.0in,height=5.2cm]{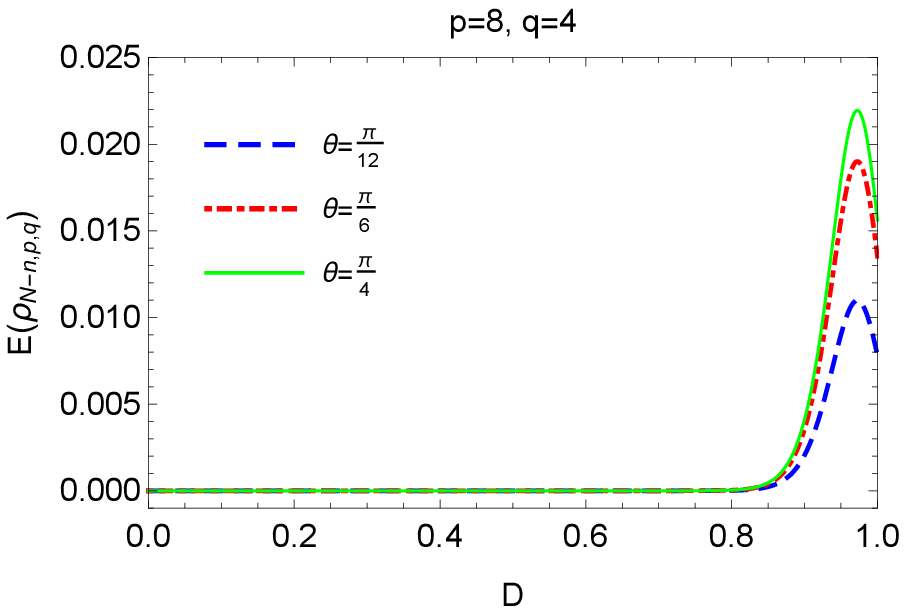}
\label{fig1a}
\end{minipage}%
\begin{minipage}[t]{0.5\linewidth}
\centering
\includegraphics[width=3.0in,height=5.2cm]{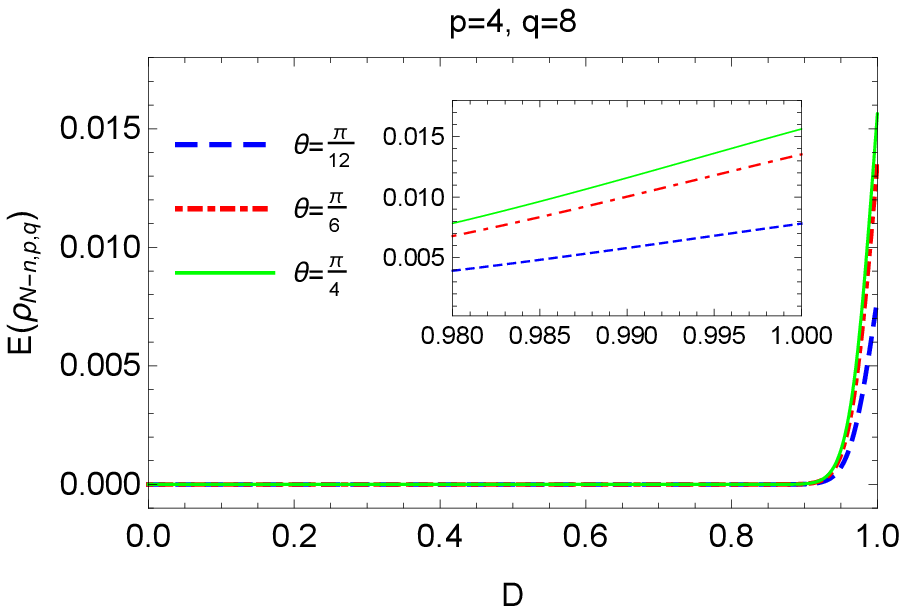}
\label{fig1c}
\end{minipage}%

\begin{minipage}[t]{0.5\linewidth}
\centering
\includegraphics[width=3.0in,height=5.2cm]{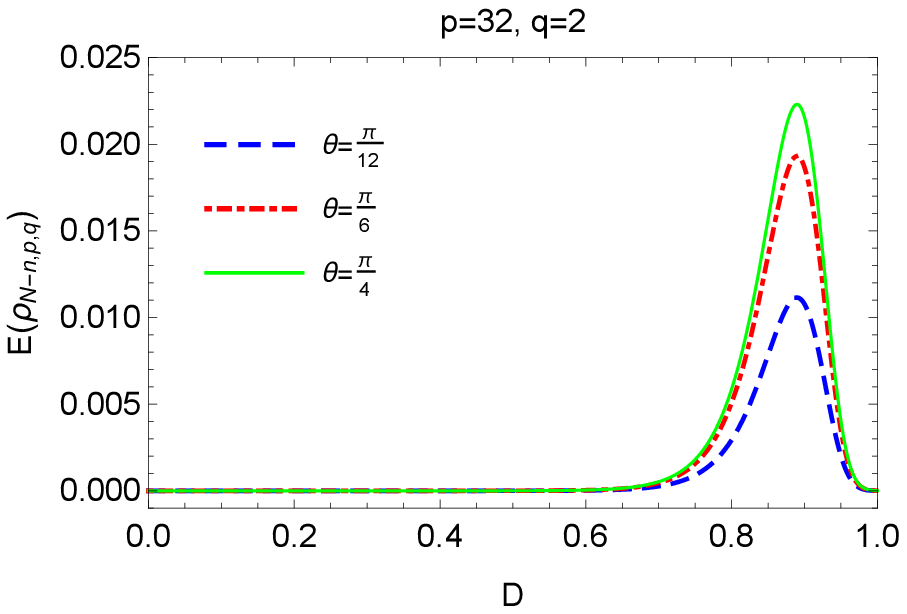}
\label{fig1a}
\end{minipage}%
\begin{minipage}[t]{0.5\linewidth}
\centering
\includegraphics[width=3.0in,height=5.2cm]{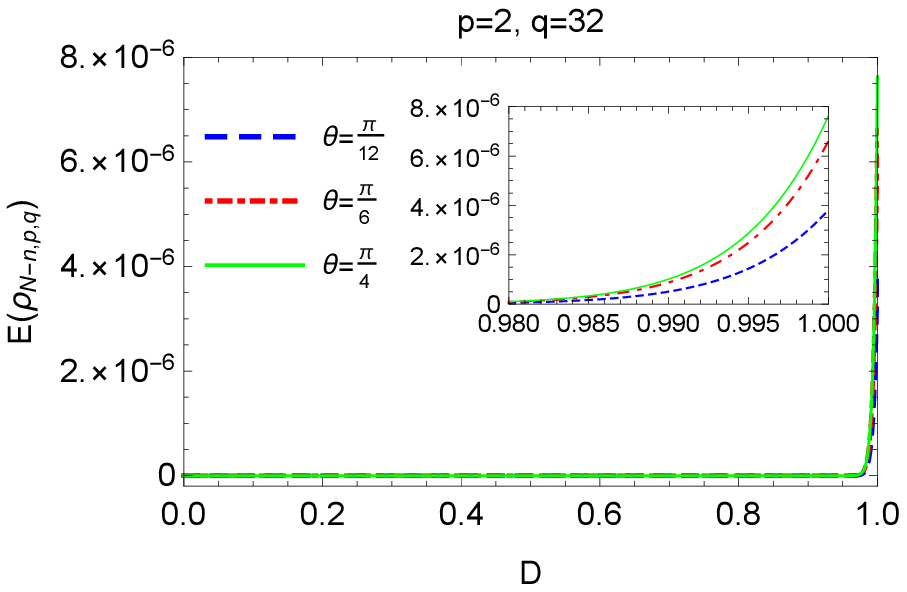}
\label{fig1c}
\end{minipage}%
\caption{The genuine N-partite entanglement $E(\rho_{N-n,p,q})$ as functions of the  dilaton $D$ of the GHS black hole for different $p$ and $q$, where $M=\omega=1$. $\theta=\frac{\pi}{12}$ (dashed blue line); $\theta=\frac{\pi}{6}$ (dotted red line).; $\theta=\frac{\pi}{4}$ (solid green line).}
\label{Fig3}
\end{figure}

In Fig.\ref{Fig3}, we plot the genuine N-partite entanglement $E(\rho_{N-n,p,q})$ as functions of the dilaton $D$ for different $p$, $q$ and $\theta$. Different from entanglement $E(\rho_{1,\ldots,N}^{\rm out})$ and $E(\rho_{1,\ldots,N}^{\rm in})$ shown in Fig.\ref{Fig1} and Fig.\ref{Fig2}, the  entanglement $E(\rho_{N-n,p,q})$ is not necessarily a monotonic function of the dilaton $D$ in
general, since it is a combination of a decreasing function $\alpha$ and an increasing function $\beta$. For $p>q$ (means $p>n/2$ and $q<n/2$), $E(\rho_{N-n,p,q})$ is non-monotonic, such as $\{p=8,q=4\}$ and $\{p=32,q=2\}$ in Fig.\ref{Fig3}. The maximum of $E(\rho_{N-n,p,q})$ is $\rm{exp}[-8\pi (M-D)\omega]=\frac{q}{p}$. Oppositely, for $p<q$, $E(\rho_{N-n,p,q})$ is monotonic, such as $\{p=4,q=8\}$ and $\{p=2,q=32\}$. It is worthwhile to point out that if $N\leq 3$, then the non-monotonic phenomenon never occurs. For example, if $N=3$, then we have three possible cases: i) $p=q=1$ and $N-n=1$; ii) $p=0$, $q=2$ and $N-n=1$; and iii) $p=2$, $q=0$ and $N-n=1$. The first two cases correspond to monotonically increasing functions, and the last case corresponds to monotonically decreasing function [which is actually the Eq.(\ref{Q18})]. The result is consistent with the previous one \cite{L48,L49,L50,L51,L52,L53,L54,L55,L56,L57}. Note that Eq.(\ref{Q27}) also has the
asymptotic value $\frac{\sin2\theta}{(\sqrt{2})^n}$  in the limit of $D\rightarrow M$.

Fig.\ref{Fig3} also shows that, under given dilaton $D$, $E(\rho_{N-n,p,q})$ increases when $\theta$ goes up from $\pi/12$ to $\pi/4$. This is because the initial state of Eq.(\ref{Q14}) has its maximal entanglement for $\pi/4$, and Fig.\ref{Fig3} just suggests that the dependency of entanglement on $\theta$ remains unchanged under the Hawking radiation. The same phenomenon also exists in  Fig.\ref{Fig1} and  Fig.\ref{Fig2} indeed.

In the above content, we have studied the gravitational effect  on  the genuine N-partite entanglement. We have seen that the physically accessible
entanglement degrades; at the same time, the physically inaccessible entanglement increases monotonically or non-monotonically, as the dilaton $D$ of the GHS black hole rises. A question naturally rises: are there any relationships between them? The answer is positive.
Through careful observation of Eq.(\ref{Q27}), we obtain some relationships between accessible and inaccessible genuine N-partite entanglement,
which are called the distributed relationships of quantum entanglement.
The first distributed relationship of quantum entanglement is
\begin{eqnarray}\label{Q28}
\sum_{p=0}^nC_n^p E^2(\rho_{N-n,p,q})&=&\sum_{p=0}^n4C_n^p\alpha^{2p}\beta^{2q}\cos^2\theta\sin^2\theta\\ \nonumber
&=&4\cos^2\theta\sin^2\theta(\alpha^{2}+\beta^{2})^n,
\end{eqnarray}
where the binomial coefficient is defined by $C_n^p=\frac{n!}{p!(n-p)!}$. Due to $\alpha^{2}+\beta^{2}=1$, we obtain
\begin{eqnarray}\label{Q29}
\sum_{p=0}^nC_n^p E^2(\rho_{N-n,p,q})=\sin^22\theta.
\end{eqnarray}
The second distributed relationship of quantum entanglement is
\begin{eqnarray}\label{Q30}
\sum_{q=0}^{\frac{n}{2}}C_{\frac{n}{2}}^q E(\rho_{N-n,p,q})=\sin2\theta,
\end{eqnarray}
where $n$ must be an even number.
Note that $\sin2\theta$ is the initial entanglement of the state Eq.(\ref{Q14}), thus Eqs.(\ref{Q29}) and (\ref{Q30}) offer the restrictive relations between physically accessible and inaccessible genuine N-partite entanglement. This means that the distributed relationships  control how quantum information travels inside and outside the event horizon of the black hole.
When one of them increases, there must be reduction of
some other elements of them. Note that when $n$ increases, the entanglement ways for $E(\rho_{N-n,p,q})$, i.e., $\sum_{p=0}^nC_n^p$ or $\sum_{q=0}^{\frac{n}{2}}C_{\frac{n}{2}}^q$ increase. However, the restrictive relations Eqs.(\ref{Q29}) and (\ref{Q30}) must be met.
Thus every entanglement $E(\rho_{N-n,p,q})$ must decrease. This is an alternative explanation of the observed phenomenon that the entanglement decreases with $n$ in Fig.\ref{Fig1} and Fig.\ref{Fig2}.

Besides the distributed relationships, we also find that the general genuine N-partite entanglement $E(\rho_{N-n,p,q})$  satisfies the following Coffman-Kundu-Wootters monogamy inequality
\begin{eqnarray}\label{Q31}
E^2(\rho_{N-n,p,q})-\sum_{i=2}^NE^2(\rho_{1,i})=\alpha^{2p}\beta^{2q}\sin^22\theta>0,
\end{eqnarray}
where $E(\rho_{1,i})$ is the bipartite entanglement between the first observer and the $i$ ($i>1$) observer. It should be emphasized that all bipartite entanglement is equal to zero  in the background of GHS dilaton black hole.
This implies that the Coffman-Kundu-Wootters monogamy inequality of N-partite is still valid in the context of black hole.

%------------------------------------------------------------------------------------------------------------------------------------------------------------------------------------------------%
\section{ Conclusions  \label{GSCDGE}}
%--------------------------------------------------------------------------------
The gravitational effect on the genuine N-partite entanglement of Dirac fields
in GHS dilaton  spacetime has been investigated. We assume that $N$ observers initially hold a $N$-mode entangled state of Dirac fields in the asymptotically flat region. Then let $n$ ($n<N$) observers hover near the event horizon of the GHS
dilaton black hole, and the rest $N-n$ observers still stay stationarily at the asymptotically flat region.
By calculating the general analytic expression that includes all physically accessible and inaccessible genuine N-partite entanglement, we have found that the physically accessible genuine N-partite entanglement appears irreversible decoherence phenomenon with the increase of the dilaton, and approach to zero for $n\rightarrow\infty$.
On the other hand, we have found that the physically inaccessible genuine N-partite entanglement
increases either monotonically or non-monotonically with the increase of the dilaton, depending on the relative numbers between the accessible and the inaccessible modes. This result is quite different from the cases of the physically inaccessible bipartite and tripartite entanglement, which change only monotonically with the increase of the dilaton \cite{L48,L49,L50,L51,L52,L53,L54,L55,L56,L57}. The maximum for the non-monotonic increase of the inaccessible entanglement is $\rm{exp}[-8\pi (M-D)\omega]=\frac{q}{p}$, i.e.,equals the ratio of the numbers between inaccessible and accessible modes.
In addition, we have found two distributed relationships  between accessible
and inaccessible genuine N-partite entanglement, and verified that the Coffman-Kundu-Wootters monogamy inequality of N-partite entanglement is still valid in the
background of dilaton black holes.

Since the notion of Hawking radiation was presented, the black hole information paradox has been attracted and puzzled scientists. Recently, new developments using holography and island paradigm have been made\cite{Yu2022,Ahn2022}.
It was shown that the construction of islands is the key to resolve the black hole information paradox. Without island, the entanglement entropy of Hawking radiation grows linearly in time and becomes divergent at late times. While taking
account of the existence of the island outside the event horizon, the entanglement entropy stops growing at late times and eventually reaches a saturation value. In our paper, we have employed the N-partite entanglement between accessible and inaccessible modes to describe the characteristic of Hawking radiation, which is actually equivalent to the description of entanglement entropy: Hawking radiation causes information traveling from the exterior to the interior of the event horizon, leading to the production of the N-partite entanglement between accessible and inaccessible modes. Accordingly, the entanglement entropy of Hawking radiation grows. Further, the description of N-partite entanglement in our work has a merit: The problem of information paradox never occurs. Because the distributed relationships for the N-partite entanglement between accessible and inaccessible modes imply that the entanglement is bounded by its initial value. The N-partite entanglement never diverges. We expect that the research can help us understand how quantum information travels inside and outside the event horizon of the black hole, and provide some new understanding of the Hawking radiation.

\begin{acknowledgments}
This work is supported by the National Natural
Science Foundation of China (Grant Nos.1217050862, 11275064), and BNUXKJC2017.	
\end{acknowledgments}

%\newpage

\end{document}